\documentclass[nofootinbib,floatfix,superscriptaddress,twocolumn]{revtex4}
\usepackage{graphicx}
\usepackage{epsfig}
\usepackage{bm}
\usepackage{comment}
\usepackage[T1]{fontenc}
\usepackage[latin9]{inputenc}
\usepackage{amssymb}
\usepackage{float}
\usepackage{amsmath}
\usepackage{dcolumn}
\usepackage{cancel}
\usepackage[colorlinks]{hyperref}
\usepackage[usenames,dvipsnames]{color}
\hypersetup{
     breaklinks=true,
    pdfstartview={FitH},    
    colorlinks=true,       
    linkcolor=blue,          
    citecolor=red,        
    filecolor=magenta,      
    urlcolor=blue,           
    anchorcolor=green,      
    linktocpage=true
}

\newcommand{\be}{\begin{equation}}
\newcommand{\ee}{\end{equation}}

\begin{document}

\title{Imprints of Barrow--Tsallis Cosmology in  Primordial Gravitational Waves}
%

\author{P. Jizba}
\email{p.jizba@fjfi.cvut.cz}
\affiliation{FNSPE,
Czech Technical University in Prague, B\v{r}ehov\'{a} 7, 115 19, Prague, Czech Republic}

\author{G.~Lambiase}
\email{lambiase@sa.infn.it}
\affiliation{Dipartimento di Fisica, Universit\`a degli Studi di Salerno, Via Giovanni Paolo II, 132 I-84084 Fisciano (SA), Italy}
\affiliation{INFN Gruppo Collegato di Salerno - Sezione di Napoli c/o Dipartimento di Fisica, Universit\`a di Salerno, Italy}

\author{G.~G.~Luciano}
\email{giuseppegaetano.luciano@udl.cat}
\affiliation{Department of Chemistry, Physics and Environmental and Soil Sciences, Escola Politecninca Superior, Universidad de Lleida, Av. Jaume
II, 69, 25001 Lleida, Spain}

\author{L.~Mastrototaro}
\email{lmastrototaro@unisa.it}
\affiliation{Dipartimento di Fisica, Universit\`a degli Studi di Salerno, Via Giovanni Paolo II, 132 I-84084 Fisciano (SA), Italy}
\affiliation{INFN Gruppo Collegato di Salerno - Sezione di Napoli c/o Dipartimento di Fisica, Universit\`a di Salerno, Italy}

\date{\today}

\begin{abstract}
Both the Barrow and Tsallis $\delta$ entropies are one-parameter generalizations of the black-hole entropy, with the same microcanonical functional form. 
The ensuing deformation is quantified by a dimensionless parameter $\Delta$, which in the case of Barrow entropy represents the anomalous dimension, while in Tsallis' case, it describes the deviation of the holographic scaling from extensivity.
Here, we utilize the gravity-thermodynamics conjecture with the Barrow--Tsallis entropy to investigate the implications of the related modified Friedmann equations on the spectrum of primordial gravitational waves. We show that, with the experimental sensitivity of the next generation of gravitational wave detectors, such as the Big Bang Observer, it will be possible to discriminate deviations from the $\Lambda$CDM model up to $\Delta\lesssim\mathcal{O}(10^{-3})$. 

 \end{abstract}

 \maketitle

\section{Introduction}
\label{Intro} 

The reconciliation of quantum mechanics and general relativity is one of the key challenges in contemporary theoretical physics. Yet although significant efforts have been made over the last few decades, a satisfactory solution to this problem has not yet been found. Instead, there are several competing approaches, each with its own technical and/or conceptual flaws (for a recent overview of the main lines of development in quantum gravity, see e.g. Ref.~\cite{Rovelli:2000aw}). 
In this situation, black holes are seen as a valuable testing ground for new ideas about quantum gravity, since it is widely agreed that they can provide a sought-after insight into how gravity should be unified both with quantum mechanics and thermodynamics.

In 2009, Tsallis proposed a thermodynamic entropy that is suitable for systems with a sub-extensive scaling of microstates, such as e.g. black holes or quantum condensed-matter systems~\cite{Tsalis:09} (see also~\cite{Tsallis:2012js,Tsab}). This, so-called $\delta$-entropy, is an entropic functional of the form:
\begin{eqnarray}
S_\delta \ = \ \eta_{\delta}\sum_i p_i \left( \log \frac{1}{p_i}\right)^{\!\!\delta}\, , \;\;\;\; \delta \ > \ 0\, ,
\label{I.1.cc}
\end{eqnarray}
where $p_i$ represent the probabilities of microstates, and the multiplicative constant $\eta_\delta$ (which is generally dependent on $\delta$) reflects the units used for measuring entropy. For uniform distribution, (\ref{I.1.cc}) reduces to its ``microcanonical'' form, namely 
\begin{eqnarray}
S_\delta \ = \ \eta_{\delta} \left(\log W \right)^{\delta} \, ,
\label{I.2.bn}
\end{eqnarray}
where $W$ is a number of available microstates. 
According to~\cite{Tsalis:09,Tsallis:2012js,Tsab}, the entropy (\ref{I.1.cc}) can be considered as a valid thermodynamic entropy for systems with the sub-extensive scaling, provided $\delta$ is appropriately chosen. To understand this, consider, for example, a black hole entropy in cosmology~\cite{HB2} or the
entanglement entropy of ground states in condensed matter theory~\cite{Srednicky,Eisert,Shor}. Such entropies exhibit the holographic-like area-law scaling, namely
\begin{eqnarray}
S \ = \ - \kappa \sum_i p_i\log p_i \  \propto \  \ L^2\, .
\label{I.3.cc}
\end{eqnarray}
Here $\kappa$ is a numerical constant which in the context of holographic thermodynamics
is commonly selected to be the  Boltzmann constant $k_B$. $L$ represents a characteristic length scale.
By the asymptotic equipartition property~\cite{Cover} the $S$ entropy in (\ref{I.3.cc}) behaves as
\begin{eqnarray}
S \ \propto \ \log W\,,
\end{eqnarray}
implying that the number of microstates $W$ (more precisely, the volume of a typical set) scales exponentially, i.e.
\begin{eqnarray}
W \ = \ f(L) \ \! \xi^{L^2}\, , \;\;\; \mbox{with} \;\;\; \xi >1 \, ,
\label{I.5.cf}
\end{eqnarray}
where $f(L)$  ``weakly'' depends on $L$, so that
\begin{eqnarray}
\lim_{L \rightarrow \infty} f(L)/L \ = \ 0\, .
\label{I.5.cff}
\end{eqnarray}
While the scaling (\ref{I.5.cf})  prevents (\ref{I.3.cc}) from being considered as full-fledged thermodynamic entropies, the entropy (\ref{I.1.cc}) may be considered as a proper thermodynamic entropy, provided a suitable scaling exponent $\delta$ is chosen~\cite{Tsallis:2012js}. 
In Ref.~\cite{Tsallis:2012js}, Tsallis argued that with an appropriate value of $\delta$, the entropy $S_\delta$ becomes extensive and, additionally,  
it preserves the desired structure of thermodynamic Legendre transforms~\cite{Tsallis:2012js,Tsab,JizLamb}.

In particular, in Cosmology, the area law in (\ref{I.3.cc}) is associated with the black hole horizon (and by extension with the area of any 
cosmological event horizon~\cite{Gibbons}). 
In this case, equations (\ref{I.2.bn}) and (\ref{I.5.cf}) imply that in the limit of large $L$, the entropy $S_{\delta}$ is given by:

\begin{eqnarray}
S_{\delta}\ = \  \gamma_{\delta} A^{\delta}\, ,
\label{S}
\end{eqnarray}
where $A$ is the horizon area, $\gamma_{\delta}$ is a $\delta$-dependent constant, which for $\delta=1$ reduces to Hawking's conventional form $\gamma=1/(4L_p^2)$ with $L_p$ being the Planck length. 
When the number of microstates scales according to (\ref{I.5.cf}), the scaling exponent $\delta$ should be 3/2 in three spatial dimensions to ensure the entropy is an extensive thermodynamic quantity~\cite{Tsallis:2012js,Tsab}.

In 2020, Barrow conjectured that the Bekenstein--Hawking entropy should follow a deformed holographic scaling at the quantum level~\cite{Barrow}. This is due to the presumed fractal-like structure of the surface of a black hole (and more generally area of any cosmological event horizon) caused by quantum fluctuations~\cite{Barrow,JalalzadehI,JalalzadehII}. So, in particular
\begin{eqnarray}
S_{{\rm{B}}} \ \propto \   L^{2+ \Delta}\, .
\label{1.7.cc}
\end{eqnarray}
Here $\Delta$ is nothing but {\em anomalous dimension} because, similar to conventional Quantum Field Theory (QFT) definition, it measures how much the scaling dimension (i.e., $2 + \Delta$)  deviates from its classical value (i.e. 2) due to quantum effects. 
Since the coupling constant in conventional quantum gravity decreases with increasing distance, the larger distance scale, the smaller the value of $\Delta$, cf. e.g.~\cite{Hooft}.  
At large scales (low energies) it may be expected that $\Delta = 0$ and one recovers the classical Bekenstein--Hawking entropy. From~(\ref{1.7.cc}) one can infer the scaling
\begin{eqnarray}
W \ = \ f(L)\ \!\xi^{L^{2+ \Delta}}\, ,  \;\;\; {\mbox{with}} \;\;\; \xi >1\, .
\label{1.8.ll}
\end{eqnarray}
%
Barrow provided  a simple ``sphereflake'' fractal model for $\Delta$, which allows only for $\Delta \in [0,1]$. 
While the Hausdorff dimension of very rough surfaces (such as sphereflake) may indeed be arbitrarily close to the embedding Euclidean dimension (i.e., max 3), the lower value of $\Delta$ may take on negative values (unlike in sphereflake) for ``spongy'' or ``porous'' surfaces. 
For example, the Hausdorff dimension of the Sierpi{\'{n}}ski carpet is approximately  $\sim 1.89$, which means that $\Delta \sim - 0.11$. Real-world porous surfaces can have even Hausdorff dimension that is close to 1, see e.g.~\cite{Tang,Xu}. 
The fact that anomalous dimensions can be negative also comes from the QFT side, where the renormalization group arguments allow for a negative $\Delta$ in various systems~\cite{Kogut}. Actually, negative logarithmic corrections to the Bekenstein--Hawking entropy, computed e.g. from the Cardy formula~\cite{Carlip}, indicate that $\Delta$ should indeed be negative.  So, more generally, one might expect $\Delta \in (-1,1]$.


By inserting Barrow's microstate scaling (\ref{1.8.ll}) into (\ref{I.2.bn}), we obtain
\begin{eqnarray}
S_{\delta} \ = \ \gamma_{\delta} \ \! A^{(1+\Delta/2)\ \! \delta}\, .
\label{I.9.cv}
\end{eqnarray}
The extensivity of the $\delta$-entropy in 3 spatial dimensions  implies that $\delta$ and $\Delta$ are not independent but they satisfy~\cite{Tsallis:2012js,Jizba:2023fkp} 
\begin{eqnarray}
\delta \ = \ \frac{3}{2+\Delta}\, . 
\label{11.cc}
\end{eqnarray}
%
%
%
%
Both the Barrow entropy and the Tsallis $\delta$-entropy have formally the same microcanonical form, which represents a one-parameter deformation of the holographic area-law.
While the Tsallis $\delta$-entropy~(\ref{I.1.cc}) represents a full entropic function that can accommodate generic microstate probabilities (as present, e.g., in canonical or grand-canonical ensembles), the Barrow entropy is only formulated in its microcanonical form, and it is implicitly assumed that it should correspond to the Shannon entanglement entropy with QFT corrections.
Since, in the following cosmological considerations, we will need only the microcanonical form of entropy --- which is the same for both entropies,
we will  refer to both~(\ref{S}) and~(\ref{1.7.cc}) collectively as the Barrow--Tsallis (BT) entropy.

The so-called BT Cosmology is an approach that uses the first and second laws of thermodynamics to modify the Friedmann cosmological equations. Due to the non-additive nature of the BT entropy, special care needs to be taken to correctly identify the integration factor of the heat one-form~\cite{JizLamb,Jizba:2023fkp}.
The standard Friedmann--Robertson--Walker (FRW) cosmological model is naturally recovered in the limit $\delta=1$ or $\Delta =0$. 

The BT Cosmology has been utilized in various contexts in recent years.
For instance, it has been applied to the Universe's evolution in extended gravity~\cite{Ghaffari:2018wks,Nunes:2015xsa,Odintsov:2023vpj,ExtAdi} and black hole physics~\cite{Abreu:2020cyv,McInnes:2020qqh,Abreu:2020rrh,Abreu:2020dyu,Abreu:2020wbz,LucianoGhaf}, where a link to other quantum gravity models has been further discussed through the study of quasinormal modes.
%
%
It has also been tested against the latest observational data, where the value of the $\Delta$ parameter in the minimalistic Barrow model has been constrained by phenomenological consistency requirements~\cite{Anagnostopoulos:2020ctz,Dabrowski:2020atl,Leon:2021wyx,Asghari:2021bqa,Jusufi:2021fek,Nojiri:2021jxf,Luciano:2022pzg,Luciano:2022ffn,DiGennaro:2022ykp,Luciano:2022viz,Ghaffari:2022skp,Luciano:2022hhy,Luciano:2023roh,Luciano:2023wtx,Denkiewicz:2023hyj,Vagnozzi:2022moj,Nojiri:2023ikl}.
%
For the case of Tsallis $\delta$-entropy, it also helped to alleviate the tension between the current bound on the PeV Dark Matter (DM) relic abundance and recent IceCube signals of high-energy neutrino events with the value of $\delta \simeq 3/2$, i.e. the value predicted by Tsallis in his 2013 paper~\cite{JizLamb}.
When one allows for scale-dependent $\Delta$ (i.e, anomalous dimension running), the BT Cosmology represents a viable cosmological model, which satisfies Big Bang Nucleosynthesis (BBN)~\cite{Jizba:2023fkp}, 
cosmic microwave background~\cite{Saridakis:2020zol} and general relativity tests. It even leads the Hubble rate that is compatible with  the current accelerated expansion of the Universe~\cite{JizLamb,Saridakis:2020zol}.

As expected, the tightest bounds come from the post-BBN epoch~\cite{Barrow:2020kug}, where standard Cosmology works well and so $\Delta$ must be very small. On the other hand, significant deviations from conventional Cosmology are expected in the early Universe phase, which is typically not directly constrained by cosmological observations. 
Direct observations of gravitational waves in the last decade by LIGO~\cite{LIGOScientific:2014pky} and VIRGO~\cite{VIRGO:2014yos} have revealed the challenging possibility of probing fundamental physics that is otherwise inaccessible through other interactions. 
Even though all the GW signals detected so far are of astrophysical origin~\cite{LIGOScientific:2016aoc,LIGOScientific:2018mvr}, primordial gravitational waves (PGWs) generated by quantum fluctuations in the inflationary period of the early Universe are likely to be recorded in the near future~\cite{Maggiore:1999vm}. 
Such signals should provide valuable information on the pre-BBN evolution history of the Universe. They would therefore be an important tool for testing non-standard cosmological scenarios in the regime where General Relativity breaks down. 

This paper aims to explore the feasibility of the BT Cosmology in precisely this high-energy regime by predicting specific signatures that would be present in the spectrum of PGWs if the BT Cosmology were a pertinent model for the early Universe. We will show that the ensuing signatures should be within the sensitivity range of the next generation of gravitational wave (GW) detectors. Some efforts in this direction already
appeared in~\cite{Feng} in  connection with the study of the stochastic gravitational wave background generated during the
first-order cosmological QCD phase transition of the early Universe.

The remainder of the work is organized as follows: in the next section, we review the modified Friedmann equations that follow from the gravity-thermodynamic application of the BT entropy.  Sec.~\ref{PGW} is devoted to understanding the propagation of GWs in the early Universe and exploring the imprints of BT Cosmology on the PGW spectrum. 
Conclusions and perspectives are finally summarized in Sec.~\ref{Conc}.

\section{Barrow--Tsallis entropy and Modified Cosmology}

In this section we briefly review the derivation of modified Friedmann equations using the gravity-thermodynamic conjecture supplied with the BT entropy instead of the area-law entropy. Here we adopt the approach of Refs.~\cite{JizLamb,Barrow:2020kug}, where the additional terms implied by the BT entropy are considered as corrections to the total energy density in the Friedmann equations. An alternative (but formally equivalent) derivation appears in~\cite{Sheykhi:2021fwh} based on the redefinition of the effective gravitational constant. 

\subsection{Barrow--Tsallis cosmology with  $\Delta = 0$}
\label{BTISC}

We start by considering a flat FRW Universe.  In this case the curvature parameter $k=0$ and the metric reads 
\be
ds^2 \ = \ \ell_{\mu\nu} \ \! dx^\mu dx^\nu \ + \ \tilde r^2\left(d\theta^2 \ + \ \sin^2\theta\, d\phi^2\right)\, ,
\label{FRW}
\ee
where $\tilde r=a(t)\hspace{0.2mm}r$, $x^0=t$, $x^1=r$, $\ell_{\mu\nu}=\mathrm{diag}\left(-1,a^2\right)$ and $a(t)$ is the time-dependent scale factor.  
We also assume that the energy content of the Universe is described by a perfect fluid. 

Because the mathematical laws governing black hole mechanics are similar to those of thermodynamics, one often formally postulates black hole thermodynamics with no reference to statistical mechanics~\cite{Bardeen}. 
This strategy was further extended by Gibbons and Hawking~\cite{Gibbons} and later by 't Hooft~\cite{Hooft-b} and Susskind~\cite{Susskind}, who showed that black hole thermodynamics is more versatile and not just confined to black holes themselves. In particular, it has been found that event horizons also have entropy and  temperature, and that it is again possible to associate thermodynamic rules with them. This picture was farther strengthened through the AdS/CFT correspondence by showing that an asymptotically AdS black-hole on the gravity side provides a thermal background for the conformal field theory (CFT) on the boundary, and that the CFT entropy corresponds to
the horizon area of the black hole in AdS~\cite{Ryu,Maldacena}.
The aforementioned connection between entropy and geometry has stimulated ongoing debate about whether cosmological systems are merely similar to thermodynamic systems, or whether they are genuine thermodynamic systems. 
Recently, Tsallis offered yet another viewpoint~\cite{Tsallis:2012js,Tsab} in which he argued that cosmological systems with horizons are true thermodynamic systems, but with non-standard state-space scaling. In such cases, he argued, one should 
replace the conventional Shannon--Gibbs entropy with an extensive but non-additive entropy, namely the $S_\delta$ entropy. This idea was further bolstered in Refs.~\cite{JizLamb,Jizba:2023fkp}, where the integrating factor for the heat one-form associated with $S_\delta$ was deduced both from Carath\'{e}odory's principle and the zeroth law of thermodynamics. The ensuing integrating factor factorizes in the product of thermal and entropic part, where the entropic part cannot be reduced to a constant (as is the case in conventional thermodynamics), due to the non-additive nature of $S_\delta$. It is interesting to note that the corresponding first law of thermodynamics, augmented with the Clausius-like heat-entropy relation, takes exactly the form predicted by the first law of black hole thermodynamics~\cite{Pad,Pad2}. This is another reason why it is convenient to treat Barrow and Tsallis entropies on the same footing.

The above gravity-thermodynamic conjecture allows to infer the cosmological equations directly from the first two laws of thermodynamics --- the first law and the generalized Clausius relation between heat and entropy applied to the Universe's  Hubble horizon of radius $\tilde r_A=1/H$, see e.g.,~\cite{Frolov:2002va,Cai:2005ra,Cai:2008gw}. Here $H=\dot a/a$ is the Hubble parameter and the overdot denotes the time derivative. By analogy with black hole thermodynamics,
one should assign a temperature to the apparent horizon. In particular, if the ensuing radius is $\tilde r_A$, then the temperature at the horizon reads~\cite{Padmanabhan:2009vy}
\be
\label{T}
T_h \ = \ \frac{1}{2\pi\tilde r_A}\, .
\ee
Here we have employed the hypothesis of a quasi-static expansion of the Universe~\cite{Luciano:2023zrx}, which ensures that the horizon temperature is 
well-defined during evolution. 
In the following, we assume that the Universe fluid is thermalized with the horizon due to long-time interactions~\cite{Padmanabhan:2009vy,Frolov:2002va,Cai:2005ra,Izquierdo:2005ku,Akbar:2006kj}. This assumption allows us to avoid the use of non-equilibrium thermodynamic techniques and the mathematical complexities they entail.

The correspondence between the black holes and the Universe can be pursued further by introducing the concept of {\em apparent horizon entropy}. In a Universe governed by the standard General Relativity, this is given by the Bekenstein--Hawking relation $S=A/A_0$, where $A=4\pi\tilde r_A^2$ is the apparent horizon surface and $A_0=4L_p^2$.

At this stage, we can utilize the first law of black-hole thermodynamics  in the form
\begin{eqnarray}
dU\ = \ T_h dS \ - \ \mathcal{W}dV\,,
\label{14c}
\end{eqnarray}
where $\mathcal{W}$  is the work density, which (similarly as, e.g. in fluid dynamics) overtakes the role of pressure.  The entropy $S$ represents the Bekenstein--Hawking area law entropy. Alternatively, when using the integrating factor for the Tsallis $S_{\delta =3/2}$ entropy, $S$ represents $S_{\delta=1}$, which is again the Bekenstein--Hawking entropy, cf. Ref.~\cite{JizLamb}.

In the FRW Universe, the energy-momentum tensor
has the perfect fluid form: $T_{\mu\nu}=(\rho+p)u_{\mu}u_{\nu}+pg_{\mu\nu}$, with $\rho$ and $p$ being the energy density and pressure, respectively. In addition, $\rho$ and $p$ are related by the continuity equation
\begin{eqnarray}
 \dot{\rho}\  = \ - 3H(\rho \ + \ p) \, , 
 \label{15.cc}
\end{eqnarray}
where $H $ is the Hubble parameter. 
The corresponding work density (which is due to the change in the apparent horizon radius) assumes the form $\mathcal{W}=- \frac{1}{2}{\mbox{Tr}} (T^{\mu\nu}) =\frac{1}{2}(\rho-p)$ where $ {\mbox{Tr}} (T^{\mu\nu}) =  T^{\alpha \beta} h_{\alpha\beta}$.

As the Universe evolves over an infinitesimally small time interval $dt$, the increase in the internal energy $dU$ due to the change in the apparent horizon volume corresponds to the decrease in the total energy content $E$ of the Universe inside the volume, i.e. $dU =-dE$. 
%
%
%
With this, we can rewrite  (\ref{14c}) as 
\begin{eqnarray}
dE \ = \ - T_hdS \ + \ \mathcal{W}dV\, .
\label{16.cd}
\end{eqnarray}
%
By employing the relation $E=\rho V$,
where $V=\frac{4\pi}{3}\tilde{r}_{A}^{3}$ is the apparent horizon volume, and the continuity equation (\ref{15.cc}), we obtain that 
\begin{eqnarray}
T_hdS \ = \ A\left(\rho+p\right)H \tilde r_A dt \ - \ \frac{1}{2 }  A\left(\rho+p\right) \dot {\tilde r}_A dt\, ,
\end{eqnarray}
where $\dot {\tilde r}_A \equiv d{\tilde r}_A/dt$.   By assuming that the apparent horizon radius is (almost) fixed, so that  $\dot {\tilde r}_A\ll 2H\tilde r_A$, cf.~\cite{CaiKim,Sheykhi}, we obtain
\be
T_hdS \ = \ A\left(\rho+p\right)H \tilde r_A dt\, .
\ee
%
At this stage, we can easily obtain the Friedmann equations of the standard General Relativity by using the formula (\ref{T}) for the temperature $T_h$ together with the entropy-area law for $S$. In particular, we get
%
\begin{eqnarray}
\label{F1}
&&\dot H_{GR} \ = \ -4\pi G\left(\rho+p\right)\,,\\[2mm]
&&H_{GR}^2 \ = \ \frac{8\pi G}{3}\rho\,,
\label{F2}
\end{eqnarray}
where  $G = L_p^2$  is the Newtonian constant of gravitation.

%
It is worth of stressing that only the microcanonical form of the Tsallis and Barrow entropies was used to derive the Friedmann equations. Since on the level of the first law of thermodynamics 
they enter in the same way,  the Tsallis and Barrow entropies leads to the same  Friedmann equations. This fact holds also when we start to consider $\Delta \neq 0$.

\subsection{Barrow--Tsallis Cosmology with  $\Delta \neq 0$}
\label{BTIISC}
%
We have just seen that the first law of black-hole thermodynamics provides conventional cosmological equations of GR for a homogeneous and isotropic Universe.   
The question naturally arises as to how the cosmological equations for a homogeneous and isotropic Universe 
will change if a different (non-holographic) state-space scaling is considered.
%
In the case of the BT model with entropy~\eqref{I.9.cv}, one can follow the same steps as in the case of $\Delta = 0$. This will result in modified cosmological equations~\cite{Saridakis:2020lrg,Leon:2021wyx}
\begin{eqnarray}
\label{FM1}
&&\mbox{\hspace{-11mm}}\dot{H} \ = \ -4\pi G\left(\rho_m+p_m+\rho_r+p_r+\rho_{_\mathrm{DE}}+p_{_\mathrm{DE}}\right),\\[2mm]
&&\mbox{\hspace{-11mm}}H^2 \ = \ \frac{8\pi G}{3}\left(\rho_m+\rho_r+\rho_{_\mathrm{DE}}\right) ,
\label{FM2}
\end{eqnarray}
where, for later convenience, we have separated the contributions of {\em matter} (baryons plus dark matter), {\em radiation} and {\em dark energy}. Notice that the latter term is an \emph{effective} component, since dark energy does not appear explicitly in Eqs.~\eqref{FM1} and~\eqref{FM2}, but is introduced by the presence of $\Delta \neq 0$ in BT entropy.

In the following we will assume the equation of state ~$p_m = 0$, which corresponds to pressureless dust matter. The energy density and pressure of the effective dark energy are then given by~\cite{Leon:2021wyx}
\begin{eqnarray}
&&\mbox{\hspace{-5mm}}\rho_{_\mathrm{DE}}\ = \ \frac{3}{8\pi G}\left\{
\frac{\Lambda}{3}+H^2\left[1-\frac{\beta\left(\Delta+2\right)}{2-\Delta}H^{-\Delta}
\right]
\right\},~~~~~~\\[2mm]
\nonumber
&&\mbox{\hspace{-5mm}}p_{_\mathrm{DE}} \ = \ -\frac{1}{8\pi G}\left\{\Lambda+2\dot H\left[1-\beta\left(1+\frac{\Delta}{2}\right)H^{-\Delta}\right]\right.\\[2mm]
&&\mbox{\hspace{8mm}}+ \ \left.3H^2\left[1-\frac{\beta\left(2+\Delta\right)}{2-\Delta}H^{-\Delta}
\right]
 \right\},
\end{eqnarray}
where $\Lambda=4CG\left(4\pi\right)^{\Delta/2}$, $C$ is an integration constant and $\beta=\left(\pi/G\right)^{\Delta/2}$ for $A_0=4G$.

Let us now define the fractional energy density parameters
\begin{eqnarray}
\label{Omegr}
&&\hspace{-6mm}\Omega_i=\frac{8\pi G\rho_i}{3H^2}\,,\quad \Omega_\Lambda=\frac{\Lambda}{3H^2}\,,\\[2mm]
&&\hspace{-6mm}\Omega_{_\mathrm{DE}} \ = \ \frac{8\pi G\rho_{_\mathrm{DE}}}{3H^2}\ = \  1+\Omega_\Lambda-\frac{\beta\left(\Delta+2\right)H^{-\Delta}}{2-\Delta}\,,~~~~
\end{eqnarray}
where $i=m,r$ refers to matter and radiation, respectively. From Eq.~\eqref{FM2} follows that the dimensionless cosmological equation takes the simple form:
\be
\Omega_m \ + \ \Omega_r \ + \ \Omega_\Lambda \ = \ \frac{\beta\left(\Delta+2\right)H^{-\Delta}}{2-\Delta}\,,
\label{27.cv}
\ee
which can be equivalently written as
\be
\Omega_m \ + \ \Omega_r \ + \ \Omega_{DE} \ = \ 1\,.
\ee
By assuming that the matter and radiation components evolve in the conventional way, i.e.
\be
\rho_m \ = \ \rho_{m0}\left(z+1\right)^3\,,\quad \rho_r \ = \ \rho_{r0}\left(z+1\right)^4\,,
\ee
where the redshift variable $z(t) = a^{-1}(t) -1$, the dimensionless cosmological equation~(\ref{27.cv}) can be recast to
\begin{eqnarray}
E(z)\!&\equiv&\!\frac{H(z)}{H_0}\nonumber \\[2mm]
&=&\!\left[\Omega_{m0}\left(z+1\right)^3 \ + \
\Omega_{r0}\left(z+1\right)^4 \ + \ \Omega_{\Lambda 0}\right]^{\frac{1}{2-\Delta}}\nonumber \\[2mm]
&\times&\! \left[\frac{\bar{\beta}\left(2-\Delta\right)}{2+\Delta}\right]^{\frac{1}{2-\Delta}}.
\label{NMFE}
\end{eqnarray}
Here $\bar{\beta}=H_0^\Delta(G/\pi)^{\Delta/2}$ and the subscript ``$0$'' denotes quantities at the present cosmological time, where $z=0$. To evaluate~(\ref{NMFE}), we employ the values: $\Omega_{m0}=0.044$, $\Omega_{r0}=2.469\times 10^{-5}h^{-2}\left(1+0.2271N_{\mathrm{eff}}\right)$, where $h=H_0/100\,\mathrm{kms}^{-1}\mathrm{Mpc}^{-1}\simeq0.71$ is the reduced current Hubble constant and $N_{\mathrm{eff}}=3.04$ the number of relativistic species~\cite{Wilkinson}. The value of $\Omega_{\Lambda0}$ can be deduced from the flatness condition $E(0)=1$, which from~(\ref{NMFE}) reads
\begin{equation}
    \Omega_{\Lambda 0} \ = \ \left(\frac{2+\Delta}{2-\Delta}\right)\frac{1}{\bar\beta} \ - \ \Omega_{m0} \ - \ \Omega_{r0} \,.
\end{equation}
We also remind that the $z$-dependence in Eq.~\eqref{NMFE} can be converted to 
a temperature-dependence through the relation $Ta(t) = T_0 $, where $T_0\simeq 3\,\mathrm{K}$ is the average temperature of the observable Universe at present time.  The latter is true for any arbitrary $\Delta$, see e.g.,~\cite{Jizba:2023fkp}, and it might be equivalently written as
\begin{equation}
1+z \ = \ \frac{T}{T_0}
\,,
\end{equation}

The modified Friedmann equation~\eqref{NMFE} will be the key ingredient in our subsequent analysis.
In passing, one can easily check that the standard General Relativity is restored for $\Delta=0$ or equivalently for  $\delta =3/2$.


\section{Primordial Gravitational Waves}
\label{PGW}

Primordial Gravitational Waves are believed to carry the imprint of quantum fluctuations and prospective phase transitions that occurred during the inflationary period of the early Universe~\cite{Maggiore:1999vm}. Detection of such signatures would be of the utmost importance, as it would allow us to probe the history of the Universe prior to BBN (e.g., the reheating phase, the hadron and quark epochs, the early non-standard matter-dominated or kination-dominated phases, etc.)~and estimate the ensuing implications this would impose on the Standard Model of particle physics. Since General Relativity is expected to break down in the UV regime due to quantum corrections, the pre-BBN epoch provides an ideal testing ground for modified gravity theories of the early Universe. 
When the production of GWs happens in scenarios richer than the standard single-field slow-roll, the GW signal becomes potentially detectable also on scales smaller than the Cosmic Microwave Background (CMB). 
In this section, we compute the PGW spectrum in the BT Cosmology and compare it with predictions of the standard cosmological scenario. Expected measurable consequences will be discussed in some detail.

\subsection{PGW in Standard Cosmology}
\label{PGW SM}

Let us start with standard Cosmology that will serve us to 
define the relic  density of PGW, $\Omega_{\mathrm{GW}}$. This, in turn, will also help us to set up the notation. We shall focus on  tensor perturbations on an isotropic, uniform and flat background space-time.
%
In this case, one can assume that the tensor perturbations $h_{00} =  h_{0i} = 0$. Here and throughout, we will be working in the transverse traceless
($TT$) gauge, where $\partial^i h_{ij}=0$ and $h^i_i=0$. The dynamics of the tensor perturbation in first-order perturbation theory is then given by~\cite{Watanabe:2006qe}
\be
\label{hdyn}
\ddot h_{ij}\ + \ 3H\dot h_{ij} \ - \ \frac{\nabla^2}{a^2}h_{ij}\ = \ 16\pi G\hspace{0.3mm} \Pi_{ij}^{TT}\,,
\ee
where $\Pi_{ij}^{TT}$ is the $TT$ anisotropic part of the stress tensor
\be
\Pi_{ij} \ = \ \frac{T_{ij}-p\hspace{0.3mm} g_{ij}}{a^2}\,,
\label{TPE}
\ee
with $T_{ij}, g_{ij}$ and $p$ being the stress-energy tensor, the metric tensor and the background pressure, respectively. Note that for a perfect fluid, $\Pi_{ij}=0$, but this would not be true in general.

In the following analysis, it will be sufficient to consider only perturbations in the high-frequency range (in particular, we neglect contributions at much lower frequencies than $10^{-10}\,\mathrm{Hz}$ that arise from the free flow of neutrinos and photons~\cite{Weinberg:2003ur}). To solve~\eqref{hdyn}, it is convenient to work in the Fourier space, where we have~\cite{Watanabe:2006qe}
\be
h_{ij}(t,\vec{x}) \ = \ \sum_{\lambda}\int\frac{d^3k}{\left(2\pi\right)^3}\hspace{0.2mm}h^\lambda(t,\vec{k})\hspace{0.2mm}\epsilon^\lambda_{ij}(\vec{k})\hspace{0.2mm}e^{i\vec{k}\cdot\vec{x}}\,.
\ee
Here,  $\epsilon^\lambda$ is the spin-2
polarization tensor obeying the orthonormality condition $\sum_{ij}\epsilon^\lambda_{ij}\epsilon^{\lambda'*}_{ij}=2\delta^{\lambda\lambda'}$ and $\lambda=+,\times$ are the two independent wave-polarizations.

The tensor perturbation $h^\lambda(t,\vec{k})$ can be factorized as
\be
h^\lambda(t,\vec{k}) \ = \ h_{\mathrm{prim}}^\lambda(\vec{k})X(t,k)\,,
\ee
where $k = |\vec{k}|$, $X(t,k)$ denotes the transfer function, which accounts for the time evolution of the perturbation, and $h_{\mathrm{prim}}^\lambda$ the amplitude of the primordial tensor perturbations. With the above parameterization, the tensor power spectrum takes the form~\cite{Bernal:2020ywq}
\be
\mathcal{P}_T(k) \ = \ \frac{k^3}{\pi^2}\sum_\lambda\Big|h^\lambda_{\mathrm{prim}}(\vec k)\Big|^2 \ = \ \frac{2}{\pi^2}\hspace{0.3mm}G\hspace{0.3mm} H^2\Big|_{k=aH}\,.
\ee
In turn, Eq.~\eqref{hdyn} acquires the form of  a damped harmonic oscillator-like equation
\be
X'' \ + \ 2\hspace{0.2mm}\frac{a'}{a}X' \ + \ k^2X \ = \ 0\,,
\ee
where the prime indicates derivative with respect to the conformal time $\tau$, such that $d\tau=dt/a$. For later convenience we observe that in a flat Universe, if the fluid is the dominant form of matter, we have $a(\tau)\propto \tau^{\frac{2}{1+3w}}$, which allows to cast the damping term as
\be
\label{damp}
2\hspace{0.2mm}\frac{a'}{a} \ = \ \frac{4}{\tau\left(1+3w\right)}\,,
\ee
where $w$ represents the equation of state (EoS) parameter of the fluid. 

Now, the relic density of PGW from first-order tensor perturbation in the standard model of Cosmology reads~\cite{Bernal:2020ywq,Watanabe:2006qe}
\begin{eqnarray}
\nonumber
\Omega_{\mathrm{GW}}
(\tau,k)&=&\frac{[X'(\tau,k)]^2}{12a^2(\tau)H^2(\tau)}\,\mathcal{P}_T(k)\\[2mm]
&\simeq&\left[\frac{a_{\mathrm{hc}}}{a(\tau)}\right]^4\left[\frac{H_{\mathrm{hc}}}{H(\tau)}\right]^2\frac{\mathcal{P}_T(k)}{24}\,,
\label{Ttps}
\end{eqnarray}
where on the second line we have averaged over periods of oscillations, which implies 
\be
X'(\tau,k)\ \simeq \ k\hspace{0.2mm} X(\tau,k)\ \simeq \  \frac{k\hspace{0.3mm} a_{\mathrm{hc}}}{\sqrt{2}a(\tau)}\ \simeq \
\frac{a^2_{\mathrm{hc}}\hspace{0.3mm}H_{\mathrm{hc}}}{\sqrt{2}a(\tau)}\,,
\ee
with $k=2\pi f=a_{\mathrm{hc}}H_{\mathrm{hc}}$ at the horizon crossing. 

Denoting, as usual, by $h$ the dimensionless Hubble constant, the PGW relic density at present time is found to be
\begin{eqnarray}
&&\mbox{\hspace{-10mm}}\Omega_{\mathrm{GW}}(\tau_0,k)h^2 \nonumber \\[2mm]
&&
\simeq\left[\frac{g_*(T_{\mathrm{hc}})}{2}\right]\left[\frac{g_{*s}(T_0)}{g_{*s}(T_{\mathrm{hc}})}\right]^{4/3}\frac{\mathcal{P}_T(k)\Omega_{r}(T_0)h^2}{24}\,,
\label{Spt}
\end{eqnarray}
where we have indicated by $g_*(T)$ and $g_{*s}(T)$ the effective numbers of relativistic degrees of freedom that contribute to the radiation energy
density $\rho$ and entropy density $s$, respectively, i.e.
\begin{eqnarray}
\rho_r&=&\frac{\pi^2}{30}g_*(T)T^4\,,\\[2mm]
s_r&=&\frac{2\pi^2}{45}g_{*s}(T)T^3\,.
\end{eqnarray}
Additionally, $\Omega_r$ denotes the 
dimensionless radiation energy density as defined in Eq.~\eqref{Omegr}.

\begin{figure}[t]
\begin{center}
\hspace{-1mm}\includegraphics[width=8.7cm]{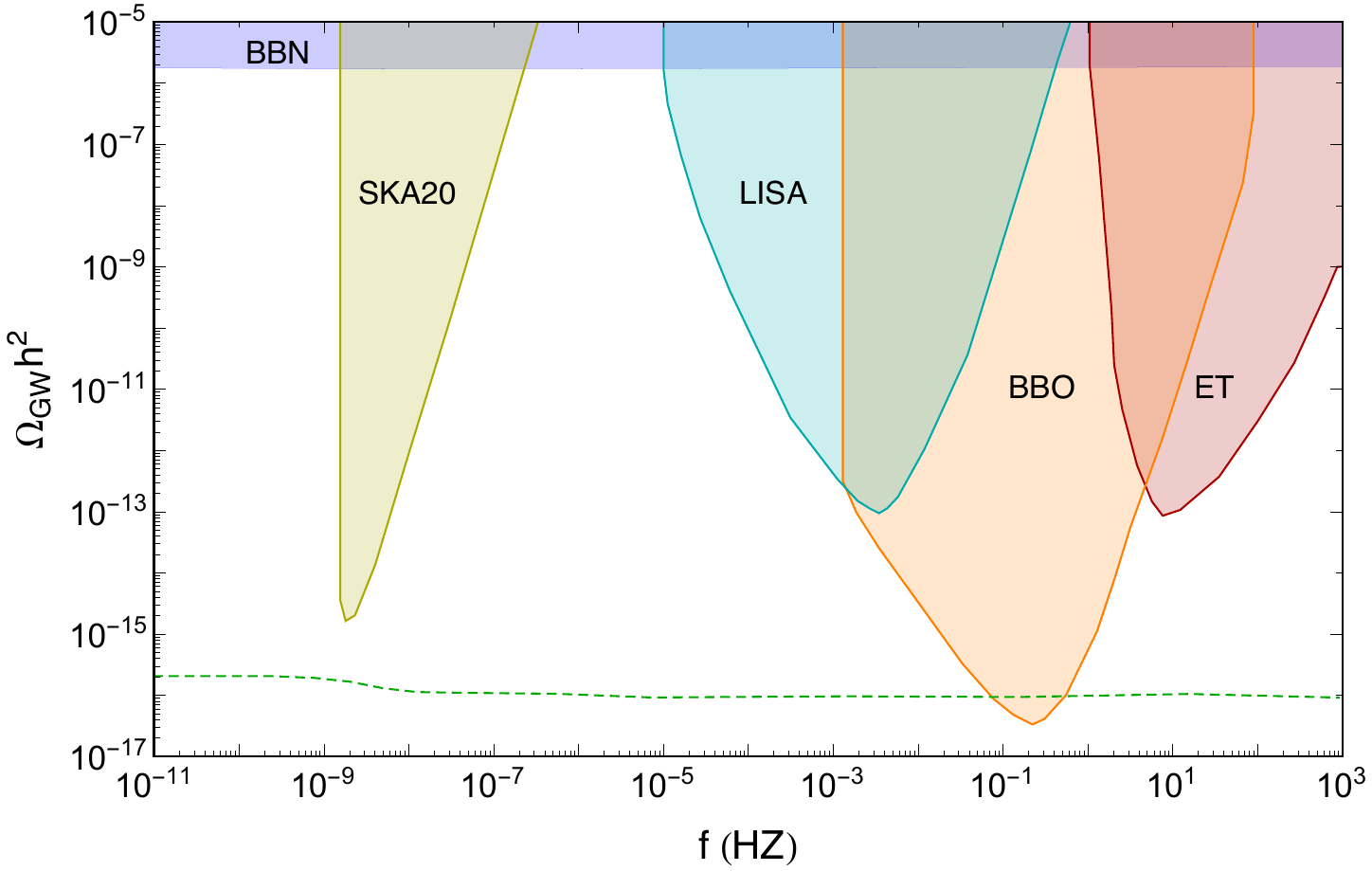}
\caption{Plot of the PGW spectrum versus the frequency $f$ for $n_T=0$ and $A_S\simeq2.1\times10^{-9}$, as predicted by the standard Cosmology (dashed green line). The colored regions delineate the projected sensitivities for several GW observatories~\cite{Breitbach:2018ddu}.}
\label{Fig1}
\end{center}
\end{figure}

The scale dependence of the tensor power spectrum is defined by
\be
\mathcal{P}_T(k) \ = \ A_T\left(\frac{k}{\tilde k}\right)^{n_T}\,,
\ee
where $n_T$ is the tensor spectral index and $\tilde k=0.05\,\mathrm{Mpc}^{-1}$ is a characteristic wave number scale. The
amplitude of the tensor perturbation $A_T$ is related to the scalar perturbation amplitude $A_S$ by $A_T=r A_S$, with $r$ being the tensor-to-scalar ratio. 

The spectrum~\eqref{Spt} is plotted against the frequency $f$ in Fig.~\ref{Fig1} (dashed green line) assuming a scale-invariant primordial
tensor spectrum (i.e. $n_T=0$) and consistently with the Planck observational constraint  $A_S\simeq2.1\times10^{-9}$ at the CMB scale~\cite{ConPlanck}. The colored regions  delineate the projected sensitivities for several GW observatories~\cite{Breitbach:2018ddu} and, in particular, constraints from LISA interferometer~\cite{LISA:2017pwj}, Einstein Telescope (ET) detector~\cite{Sathyaprakash:2012jk}, the successor Big Bang Observer (BBO)~\cite{Crowder:2005nr} and Square Kilometre Array (SKA) telescope~\cite{Janssen:2014dka}. Additionally, the BBN bound arises from the constraint on the effective number of neutrinos~\cite{Boyle:2007zx,Stewart:2007fu}.

\subsection{PGW in Barrow--Tsallis Cosmology}
\label{PGW Bar}

Now we will examine how BT entropy affects the PGW spectrum. To do this, 
we will employ the solution to the modified Friedmann equation~\eqref{NMFE}.
In fact, from Eq.~\eqref{Ttps}, it is possible to write
\begin{eqnarray}
 \Omega_{\mathrm{GW}}(\tau,k)\!&\simeq& \!\left[\frac{a_{\mathrm{hc}}}{a(\tau)}\right]^4\left[\frac{H_{\mathrm{hc}}}{H_{\mathrm{GR}}(\tau)}\right]^2\left[\frac{H_{\mathrm{GR}}(\tau)}{H(\tau)}\right]^2\frac{\mathcal{P}_T(k)}{24} \nonumber \\[2mm]
&=&\!\Omega^{\mathrm{GR}}_{\mathrm{GW}}(\tau,k)\left[\frac{H_{\mathrm{GR}}(\tau)}{H(\tau)}\right]^2\left[ \frac{a_{\mathrm{hc}}}{a_{\mathrm{hc}}^{\mathrm{GR}}}\right]^4\nonumber \\[2mm]
&&\times \ \left[ \frac{a^{\mathrm{GR}}(\tau)}{a(\tau)}\right]^4\left[ \frac{H_{\mathrm{hc}}}{H_{\mathrm{hc}}^{\mathrm{GR}}}\right]^2   \,, 
\label{eq:PGWBar0}
\end{eqnarray}
where the subscript/superscript ``GR'' denotes the quantities as they appear in conventional General Relativity, so for instance,   $\Omega^{\mathrm{GR}}_{\mathrm{GW}}(\tau,k)$ is the PGW relic density as predicted by conventional General Relativity, i.e., it agrees with Eq.~(\ref{Ttps}). 
%
%

Due to the fact that $E(z)=1$ for $z=0$, as seen from Eq.~\eqref{NMFE}, we can finally write that
\begin{equation}
 \Omega_{\mathrm{GW}}(\tau_0,k)
\ \simeq \ \Omega^{\mathrm{GR}}_{\mathrm{GW}}(\tau_0,k)\left[ \frac{a_{\mathrm{hc}}}{a_{\mathrm{hc}}^{\mathrm{GR}}}\right]^4\left[ \frac{H_{\mathrm{hc}}}{H_{\mathrm{hc}}^{\mathrm{GR}}}\right]^2   \,. \label{eq:PGWBar} 
\end{equation}
%
%

\begin{figure}
\begin{center}
\hspace{-1mm}\includegraphics[width=8.7 cm]{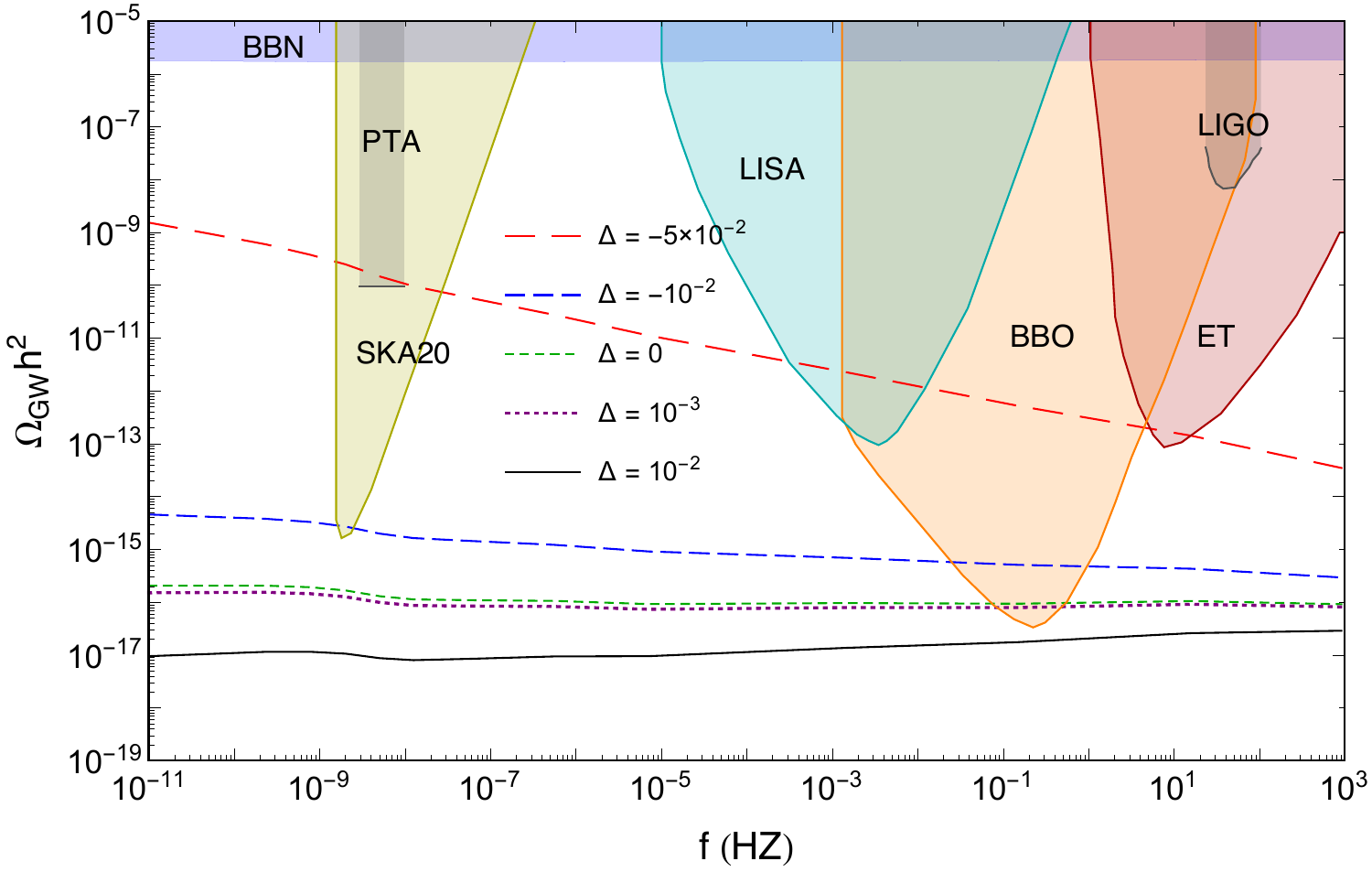}
\caption{Plot of the PGW spectrum versus the frequency $f$ for $n_T=0$ and $A_S\simeq2.1\times10^{-9}$, for different value of $\Delta$. The colored regions are the same as in Fig.~\ref{Fig1}.  In gray we show the regions excluded by  PTA~\cite{KAGRA:2021kbb} and LIGO~\cite{Shannon:2015ect}.}
\label{Fig2}
\end{center}
\end{figure}

In Fig.~\ref{Fig2} we depict the PGW spectrum $ \Omega_{\mathrm{GW}}$ as a function of the frequency $f$ for different values of $\Delta$ evaluated using Eq.~\eqref{eq:PGWBar}. For $\Delta > 0$ (i.e., the strict Barrow case), it is possible to see how, at low frequencies, the gravitational wave spectrum is reduced as $\Delta$ increases. 
In particular, with the experimental sensitivity of the BBO, if PGWs are detected, it will be possible to constrain $\Delta$ up to $\Delta\lesssim\mathcal{O}(10^{{-3}})$. On the other hand, if no signatures of PGWs are observed, this would provide further evidence that GR and the standard cosmological inflationary model need to be substantially modified for phenomenological consistency. From Barrow's point of view, the needed corrections
could be codified in the form of a modified entropy-area law with  $\Delta\gtrsim\mathcal{O}(10^{{-3}})$. From the vantage point of Tsallis, this will mean that the number of quantum microstates describing a black hole differs from (would be higher than in)  the (semi-)classical situation.

Conversely, for $\Delta < 0$ (i.e., the generalized Barrow case, corresponding to a lower number of black hole microstates in Tsallis' picture), the spectrum would be increased compared to GR. 
If this were the case in Nature, signatures of PGWs could likely be detected 
by ET, LISA and/or SKA20 (besides BBO), even for frequencies below $10^3\,\mathrm{Hz}$.
Interestingly enough, it is definitely possible to exclude in this case $\Delta\lesssim-5\times10^{-2}$, using the Pulsar Timing Array (PTA) data.  


For comparison with literature, we would like to mention that different upper bounds on $\Delta$ have been set so far, the most stringent being $\Delta\lesssim10^{-4}$ from BBN~\cite{Barrow:2020kug} and inflation~\cite{Luciano:2023roh} measurements.
On the other hand, the scenario with a fractal index bounded from below seems to be favoured by the full dynamical and geometrical data set in Cosmology~\cite{Denkiewicz:2023hyj}. Furthermore, the possibility of an energy scale-dependent behavior for fractal corrections has been explored in the recent work~\cite{Var}, based on considerations from quantum field theory and quantum gravity under renormalization. 
%

\section{Conclusions and Discussion}
\label{Conc}

The Barrow--Tsallis entropy is a one-parameter deformation of the semiclassical Bekenstein--Hawking holographic entropy, arising either from quantum gravitational corrections to the black hole horizon surface (Barrow case) or from the requirement of thermodynamic extensivity (Tsallis case). Despite their different motivations, the deformation parameters $\Delta$ and $\delta$ of the Barrow and Tsallis entropies, respectively, are directly related by the formula (\ref{11.cc}). 
In addition, both entropies lead to the same first law of thermodynamics~\cite{JizLamb,Jizba:2023fkp} and thus provide identical generalized Friedmann equations, leading to the same generalized Cosmology --- BT Cosmology.
In this work, we have studied the Hubble expansion of the Universe in the context of the BT Cosmology, and the resulting effects on the relic density of the PGW spectrum.
We have found that 
this spectrum is reduced with respect to GR as $\Delta > 0$ increases, in such a way that, 
if PGWs are going to be detected with the  experimental sensitivity of BBO, 
we are able to constrain the fractal index up to $\Delta\lesssim10^{-3}$. On the other hand, for $\Delta < 0$ the spectrum increases compared to GR and the resulting PGWs should be detectable with the experimental sensitivities of SKA, LISA, BBO and ET observatories. Interestingly, the PTA data exclude values of $\Delta\lesssim-5\times10^{-2}$ that are otherwise allowed in the $S_{\delta}$ entropy based Cosmology. This latter fact restricts $\delta$ to values $\lesssim 1.54$, which is in agreement with recent bounds on $\delta$ coming, e.g. from BBN or the Relic Abundance of Cold Dark Matter Particles~\cite{Jizba:2023fkp}.
It should be noted in passing that $\Delta <0$ is also favored by explicit calculations of the Bekenstein--Hawking entropy,
e.g. from the Cardy formula. In this case, the lowest order
correction to the classical scaling dimension is indeed negative~\cite{Carlip}.

Some other aspects remain to be investigated: first, we want to further explore the predictions of BT Cosmology in the very early Universe, where quantum gravity corrections are expected to be more relevant. Preliminary studies in this direction already appeared in~\cite{Luciano:2023roh} and~\cite{Feng} in connection with the inflation and the stochastic GW background generated from
cosmological QCD phase transition, respectively. 
In addition, it would be suggestive to develop a more fundamental Hamiltonian formulation of the Barrow ``sphereflake'' fractal model, which could serve as a step towards a better understanding of quantum gravity. 
Work is already underway and will be elaborated elsewhere.

\vspace{2mm}
\acknowledgements
PJ was in part supported by the Ministry of education grant M\v{S}MT RVO 14000. GGL acknowledges the Spanish ``Ministerio de Universidades''
for the awarded Maria Zambrano fellowship and funding received
from the European Union - NextGenerationEU.
He is also grateful for participation to the  LISA Cosmology Working group.
GL and LM thank MUR for support. GL thanks INFN for support. 
GL and GGL acknowledge the participation to the COST
Action CA18108  ``Quantum Gravity Phenomenology in the Multimessenger Approach''. 


\end{document}